\documentclass[prd,aps,superscriptaddress,twocolumn]{revtex4}
\usepackage{amsmath,amssymb}

\usepackage{graphicx}

\newcommand{\be}{\begin{eqnarray}}
\newcommand{\ee}{\end{eqnarray}}

\def\ben{\begin{equation}}
\def\een{\end{equation}}
\def\bena{\begin{eqnarray}}
\def\eena{\end{eqnarray}}

\begin{document}

\title{Spontaneous localization on a brane via a gravitational mechanism}

\author{Cristiano Germani}
\email{cristiano.germani@lmu.de}
\affiliation{Arnold Sommerfeld Center, Ludwig-Maximilians-University, Theresienstr. 37, 80333 Muenchen, Germany}

\begin{abstract}
In this letter we introduce a novel way to spontaneously localize particles (including gauge bosons) and gravitons kinetic terms on a four-dimensional brane via a gravitational mechanism. The model underlying this localization mechanism can be considered as a high-energy resolution of the so-called braneworlds scenario. In particular, we show how to construct a braneworld with induced gravity without pathologies. Finally, we argue that the brane is only stable if its own mass does not exceed a critical value related to the particles-gravity couplings.
\end{abstract}

\pacs{...}

\maketitle

\section{Introduction}

Gravity is the weakest force in Nature. It is so weak that the weakest particle physics force, the weak interaction, is still $32$ order of magnitude stronger than gravity. This unnatural mismatch generates the so called hierarchy problem. In particular, because of it, the Higgs boson mass must be incredibly fine tuned in order to avoid large corrections due to the Planck scale cut-off and spoil all the successes of the Standard Model of particle physics.

Obviously, in order to solve this problem, new physics should appear at least at the TeV scales where the Higgs mass starts to run consistently. A possibility is that this new physics lower enormously the gravity cut-off from the Planck to the TeV scales. This philosophy can be elegantly realized in braneworlds scenarios \cite{br,rs}. There, our Universe is a four-dimensional hypersurface (a brane), which is embedded in one or more extra-dimensions. In this case a much stronger extra-dimensional gravity looks weaker on the brane because of gravity ``leaking'' in the extra-dimensions. However, in order to have a phenomenologically viable model, all Standard Model particles (or any extension of it that we will shortly denote as SM) should mainly propagate on the brane at low energies. Moreover, gravity itself should be also localized on the brane in order not to clash with Newtonian experiments.

In the context of string theory, (stringy) gauge interactions can naturally propagate on fundamental D-branes. However, from the point of view of effective field theories and for a phenomenological study of the SM in higher dimensions (see for example \cite{br,rs}), it is of great interest to find an effective localization mechanism of particles on domain walls (braneworlds). In this respect, many phenomenological models have been suggested. While a phenomenological realization of spin $0$ and $1/2$ localization is not too hard to find \cite{at}, the same is not true for gauge vectors (massless spin $1$) (for the latest tentative see \cite{last}). 

In \cite{ds}, Dvali and Shifman argued that the only way to localize massless vector bosons on a brane is that, for some physical reason, vector bosons are confining on the bulk (the embedding space), whereas confinement is absent on the brane. A model independent way to describe this physical fact is to postulate (let us work on one extra-dimension for simplicity) that the Lagrangian of the vector field has a localized kinetic term on the brane \cite{dg}, i.e.
\be\label{vector}
\!\!\!\!\!\!\!S_{v}=-\frac{1}{4 e_5^2}\int d^4x d\chi \sqrt{-g} \left[{\rm Tr} F_{AB} F{}^{AB}+\frac{\delta(\chi)}{m}{\rm Tr}F_{\mu\nu} F^{\mu\nu}\right],
\ee
where $F_{AB}=\partial_{[A}A_{B]}+\left[A_A,A_B\right]$ is the field strength of the (non-)abelian vector $A^A$, $\chi=0$ is the brane position, $A,B...$ are bulk coordinates, $\mu,\nu...$ are brane coordinates, $m$ is a mass scale, $e_5^2$ (of length dimension) is the gauge coupling and finally $\delta(\chi)$ is the Dirac delta function. 

The field equations are 
\be\label{1}
\nabla_A F^{AB}+\frac{\delta(\chi)}{m}\partial_{\mu} F^{\mu\nu} \delta^B_\nu=0 .
\ee
Consider now a bulk with curvature radius $l$ and a flat four dimensional brane. If we are at higher energies than the curvature radius $p^2\gg l^{-2}$ (where $p$ is the four dimensional momentum parallel to the domain wall), we can safely neglect gravity \footnote{Of course, this discussion is valid if and only if $1/\sqrt{p^2}$ is above the brane thickness $L$. Moreover, we will systematically assume that the energy scale $1/L$ is much smaller than the energy scale at which gravity becomes strongly coupled.}. In this case, for $m$ small enough, i.e. for $p^2\gg m^2$ the term proportional to the Dirac delta function decouple in (\ref{1}) and, on the brane ($\chi\rightarrow 0$) the gauge boson propagates as in four dimensions \cite{ds} 
\be
\partial_{\mu} F^{\mu\nu}=0\ \ ,\ \square_4\, A^\chi=0 .\nonumber
\ee 
More specifically, standard calculations \cite{ds} show that the four dimensional vector field propagator on the domain wall is proportional to $\frac{1}{p^2+m p}$ and therefore the propagation is four dimensional for $p^2\gg m^2$ as announced.

An observer on the brane will then see a localized massless vector $A^\mu$ and a decoupled massless scalar $A^\chi$.
The same would happen for low energies $p^2\ll l^{-2}$ in the case in which the curvature radius $l$ is much smaller than the scale $1/m$. In fact, in this case the Dirac delta function would again dominate over derivatives of the vector $A^\mu$ in the extra-dimensional direction \cite{rsds}.

In its simplicity, the Dvali-Shifman localization mechanism is very hard to realize in dimensions larger than four as based upon confining properties of Yang-Mills theories. But this is exactly where we would need it. In other words, it would be desirable to have an effective coupling of the vector boson with other degrees of freedom such that, in some limit, would reproduce (\ref{vector}). 

Suppose we can find a smoothing of the Dirac delta function of (\ref{vector}) due to some interaction of the gauge field with some other degrees of freedom. Naturally, because of diffeomorphisms invariance
\be
\delta(\chi) F_{\mu\nu}F_{\alpha\beta}\gamma^{\mu\alpha}\gamma^{\nu\beta}\rightarrow \Delta(\chi)^{MNAB}F_{MN}F_{AB},\nonumber
\ee
where $\gamma^{\mu\nu}$ is the four dimensional metric and $\Delta(\chi)^{MNAB}$ is the smoothing tensor reproducing at low energies (\ref{vector}). To obtain (\ref{vector}) in some limit, we need the combination of the two requirements: a) $\nabla_A \Delta^{A\ldots}\rightarrow 0$ and b) that the smoothing tensor is peaked only in the four-dimensional legs with a thickness comparable to the brane thickness. In other words, it seems that the smoothing tensor should ``know'' about the spacetime structure of the bulk-brane system. Gravity itself is therefore the best candidate degree of freedom to couple to the gauge theory $F_{AB}$ in order to smoothly reproduce the theory (\ref{vector}).

In this paper, we will construct a gravity interaction to gauge fields such that, in the presence of a domain wall, gauge bosons would be spontaneously localized on the bulk (but not on the brane). We will indeed find a phenomenological Lagrangian that reduces to (\ref{vector}) in the presence of a brane. The same mechanism can then be applied to any spin, including gravity. In other words, we will propose here a high-energy resolution of the so-called braneworld scenarios \cite{br,rs}.

\section{Five dimensional Lagrangians}

\subsection{Gauge Bosons} 

When talking about gravity, renormalizability is not a good criteria to construct effective field theories. Nevertheless, at least at tree level, it is advisable to consider only theories with equation of motion that are at most second order in derivatives. This requirement is crucial as it avoids possible instabilities due to ghost propagations in general backgrounds. 

We will start by considering vector fields. For simplicity we will restrict to abelian fields but for non-abelian the generalization is straightforward.

Our philosophy here is to find a confining mechanism without introducing explicitly new degree of freedom more than the graviton and the gauge boson \footnote{This is not exactly true. We will indeed implicitly introduce new degrees of freedom by requiring the presence of a domain wall in the bulk.}.

A Lagrangian quadratic in the field strength, interacting to gravity and generating only second order differential equations for both vector and metric variations is \cite{unique,vector}
\be\label{sv}
S_{v}&=&-\frac{1}{4e_5^2}\int d^4x d\chi \sqrt{-g} \Big[F_{AB} F^{AB}+\cr
&+&\frac{1}{M_v^2}\Delta_v^{AB}{}_{CD}F_{AB} F^{CD}\Big],
\ee
where
\be
\Delta_v^{AB}{}_{CD}\equiv \frac{1}{8}R^{AB}{}_{CD}-\frac{1}{2}R^{[A}{}_{[C}\delta^{B]}{}_{D]}+\frac{1}{8}R\delta^{[A}{}_{[C}\delta^{B]}{}_{D]},
\ee
and $M_v$ is a mass scale. In the theory (\ref{sv}), the generation of second order equations of motion (for both metric and field variations) is due to the fact that $\nabla_A \Delta_v^{ABCD}\equiv 0$ by Bianchi identities \cite{unique,vector}. In particular, because of that, we get the following equation of motion for the vector field
\be
\nabla_A F^{AD}+\frac{1}{M_v^2}\Delta_v^{ABCD}\nabla_{A}F_{BC}=0,
\ee
that closely resemble (\ref{1}).

The Lagrangian (\ref{sv}) is actually unique in four-dimensions \cite{unique}, however it is perhaps not in five dimensions. We leave this check for future work. In any case, we would like here to provide a mechanism for gauge localization which in its generality can be realized with the theory (\ref{sv}). Finally, we note that the sign in front of the non-minimal coupling has been chosen to avoid ghost propagation in backgrounds that do not violate energy conditions. In the following, however, we will violate them as we will be interested in domain wall solutions. This will put a bound on the mass of the domain wall, as we shall show it.

We will not consider specific details of how a brane is constructed. Indeed, we will only consider energy scales in which the brane thickness is {\it not} resolved. Specifically, the physical brane will be here simulated by a lower dimensional hypersurface. Nevertheless, we will extensively use some of the ingredients of a physical domain wall. Generically, a self-gravitating domain wall may only be constructed if dominant energy conditions are violated such that the bulk spacetime generically approaches Anti-DeSitter (AdS) at large distances from the domain wall \cite{review}. The negative energy is necessary to keep the wall ``still'', otherwise it would quickly undergo a cosmological contraction (expansion). 

If we ignore the micro structure of the domain wall (its thickness), all the above properties are well captured by a warped extra-dimensional geometry of the Randall-Sundrum type \cite{rs}
\be\label{metric}
ds^2=N^2(\chi)ds^2_4+d\chi^2,
\ee
where $ds_4^2$ is the flat metric on the brane, $N(\chi)\equiv e^{-\frac{|\chi|}{l}}$ and finally $\chi$ is the extra-dimensional coordinate (in the following we will use the notation ${}'=d/d\chi$).

The second derivative of the scale factor is $\frac{N''}{N}=\frac{1}{l^2}-2\frac{\delta(\chi)}{l}$.
Curvatures are roughly proportional to the second derivative of the warp factor, schematically ${\rm ``curvatures"}\sim -\frac{1}{l^2}+\frac{\delta(\chi)}{l}$. We then see, as already announced, that the large distance metric is of the AdS type with curvature length $l$. 

We can now plug the ansatz (\ref{metric}) into the action (\ref{sv}). After a lengthly but straightforward calculation we obtain
\be
S_v&=&-\frac{1}{4e_5^2}\int d^4x d\chi \sqrt{-g} \Big[\left(1-\frac{3}{4M_v^2 l^2}\right)F_{AB} F^{AB}+\cr
&+&\frac{\delta(\chi)}{M_v^2 l}F_{\mu\nu} F^{\mu\nu}\Big],\nonumber
\ee
this lagrangian is of the form (\ref{vector}) which is our main result. However, there is more than that. We note that the gauge coupling for the bulk theory is infinite for a critical length $l_c=\sqrt{\frac{3}{4}}\frac{1}{M_v}$ and negative for $l<l_c$.
This is due to the fact that, as explained before, the bulk geometry (AdS) violates energy conditions. Therefore, in this theory, there cannot be domain walls with energy larger than $1/l_c$. In fact, if these domain walls were produced they would be quickly destroyed by the bulk gauge (ghost) instability. Note that in this process the total kinetic term never vanishes as the boundary kinetic term is always non-zero and non-ghost like.

Let us now analyze the theory (\ref{1}). In particular we can define an effective gauge coupling and a mass scale:
$g_5^2\equiv \frac{e_5^2}{l(1-\frac{l_c^2}{l^2})}$ , $m\equiv\frac{g_5^2}{e_5^2 M_v^2}$.
The theory is then conveniently rewritten as
\be\label{2}
\!\!\!\!\!\!\!\!S_v=-\frac{1}{4 g_5^2 l}\int d^4x d\chi \sqrt{-g} \Big[F_{AB} F^{AB}+\frac{\delta(\chi)}{m}F_{\mu\nu} F^{\mu\nu}\Big].
\ee
We see that for critical branes with $l\rightarrow l_c$ ($m\rightarrow 0$ but $mg_5^2\rightarrow {\rm finite}$), the bulk kinetic term vanishes, so that the {\it decoupled} vector field looks like purely four-dimensional, no matter what scale $M_v(<\infty)$ is. This limit however, corresponds to a strong coupling whenever the vector field is coupled to some {\it bulk} fields. There, the theory cannot obviously be trusted. 

As discussed in the introduction, the four-dimensional behavior of the gauge theory is determined by the scale $l$ and $m$ \cite{rsds}. In particular, smaller is $m$ more ``four-dimensional'' the gauge theory looks like. It is clear that $m$ can be chosen to be small by choosing a small coupling mass $M_v$. What is non-trivial is that the four-dimensional behavior of the gauge theory can be made parametrically better and better by considering domain walls with curvatures approaching $l_c$, but far enough from it in order to avoid strong couplings. 

\subsubsection{Quantum corrections?}

One may ask whether this localization mechanism can be spoiled by quantum corrections. Although the full quantum analysis is very hard to perform, in the weakly coupled regime we will argue that this would not be the case. In other words, we expect that the localization mechanism presented here is robust under quantum corrections. This is mainly due to the fact that, from the four-dimensional effective field theory, our theory is just a four dimensional gauge theory coupled to a resonance, as in \cite{dg}. In this case, the running of couplings is not expected to change the structure of the effective four-dimensional theory and therefore, in turn, of the localization mechanism. Let us discuss this result from the five-dimensional point of view.

The first thing to note is that quantum corrections may only be calculated on a specific background. In this sense, the theory (\ref{sv}) is in weak coupling in the background (\ref{metric}) as long as the energy scale of the system if far below the strong coupling scale of the (perturbed) theory around the same classical background. Note that strong coupling scales are indeed background dependent whenever the background is gravitationally non-trivial (for a discussion in the context of inflation see for example \cite{GEF,linde}).

Having said that, in the background (\ref{metric}) we can split the action (\ref{sv}) into a five and four dimensional one and study separately the quantum corrections whenever $l$ is far away from $l_c$. The case of a critical brane $l\sim l_c$ is, as already pointed out, describing a strongly coupled theory and therefore cannot be analyzed with standard perturbative techniques. 

Let us start with the five dimensional contribution and switch off, for the time being, gravity fluctuations around the Randall-Sundrum background (\ref{metric}). In this case, the effective gauge coupling $g_5^2$ runs only logarithmically exactly like in four-dimensions \cite{log}. We can now discuss about the gravity contribution in the stable $l>l_c$ case. The graviton propagator is canonically normalized with the Planck scale $M_p$. Therefore, the operator $\Delta^{\alpha\beta}$ would produce a dimension six interaction scaling with the cut-off $\Lambda=(M_p M_v^2)^{1/3}$. For energy scales far below $\Lambda$ the gravity contribution to a possible run of $M_v$ would then be negligible. 

We can now study the purely (boundary) four-dimensional sector. In this case again, as for $p^2\gg m^2$ the theory is four dimensional, the effective coupling constant $g_5^2 m$ would only run logarithmically just as in five dimensions. In other words, in the localization range $m^2\ll p^2\ll \Lambda^2$, the only running coupling would be the overall $e_5$ (as in four-dimensions) in (\ref{sv}) and not $M_v$. This can also be understood by noticing that thanks to Bianchi identities, and for a rigid spacetime, the whole action (\ref{sv}) may be rewritten in terms of an effective metric containing $\Delta$. In this case only the overall coupling constant would run as in \cite{log}. We conclude then that for energy scales far below $\Lambda$ the localization mechanism presented here would not be spoiled by quantum corrections. A more complete analysis is however left for a future work.
  

\subsection{Spin $0$ and $1/2$} 

Here we extend the confining mechanism introduced for the gauge sector to the matter sector. The logic is exactly the same as before. In order to achieve localization we will non-minimally couple the kinetic term of fermions and scalar fields to the Einstein tensor ($G^{AB}$) as follows:

\paragraph{Scalars}
\be
S_s=-\frac{1}{2}\int d^4xd\chi\sqrt{-g}\left[\Delta_\phi^{AB}D_A\phi D_B\phi-V(\phi)\right].
\ee

\paragraph{Fermions}
\be
S_f=-\frac{1}{2}\int d^4xd\chi\sqrt{-g}\bar\psi\left[\Delta_\psi^{AB}\Gamma_A D_B\psi+m_\psi \psi\right],
\ee
where 
\be
\Delta_{\phi,\psi}^{AB}\equiv g^{AB}-\frac{G^{AB}}{6M_{\phi,\psi}^2},
\ee
$D_A$ is the (gauge) covariant derivative, $\Gamma^A=e^A_a\tilde\Gamma^a$ where $e^A_a$ is the f\"unf-bein of the spacetime, $\tilde\Gamma^a$ are the flat space Dirac matrices in five dimensions and finally $M_\phi$ and $M_\psi$ are mass scales. 

As proven in \cite{GEF} for scalars, the non-minimal interaction of the kinetic term with the Einstein tensor does not propagate new degree of freedom (the same can be easily proven for fermions). In fact, 
only maximum second order derivatives appear in both field and gravity equation of motion thanks to the Bianchi identities $\nabla_A \Delta_{\phi,\psi}^{AB}\equiv 0$.

By using the ansatz (\ref{metric}) we obtain
\be
S_s&=&-\frac{1}{2}\int d^4xd\chi\sqrt{-g}\Big[\left(1-\frac{1}{M_\phi^2l^2}\right)D_A\phi D^A\phi+\cr
&+&\frac{\delta(\chi)}{M_\phi^2l}D_\mu\phi D^\mu\phi-V(\phi)\Big].
\ee
and
\be\label{fermion}
S_f&=&-\frac{1}{2}\int d^4xd\chi\sqrt{-g}\bar\psi\Big[\left(1-\frac{1}{M_\psi^2l^2}\right)\Gamma^A D_A\psi\cr
&+&\frac{\delta(\chi)}{M_\psi^2l}\Gamma^\mu D_\mu\psi+m_\psi \psi\Big].
\ee
We see again that there are critical bulk curvatures $l_c^{\psi,\phi}\equiv\frac{1}{M_{\psi,\phi}}$, such that if $l<l_c^{\psi,\phi}$ we get an instability. Finally, we note that the localized fermions (\ref{fermion}) are not chiral, similarly as in the case of \cite{ds}. Chiral fermions may however be easily localized by considering $m_\psi$ to have a kink profile across the brane \cite{at}.   

\subsection{Gravity}

Gravity is already localized on a warped geometry \cite{rs}. Nevertheless, gravity self-coupling via the Gauss-Bonnet combination (that propagates only two graviton polarizations) would, exactly as before for lower spins, generate a localized graviton kinetic term (induced curvature) capturing properties of the so called Dvali-Gabadadze-Porrati (DGP) scenario \cite{dgp}. For co-dimension larger than 1, the generation of a DGP-like model from Gauss-Bonnet interaction is an {\it exact} result (see for example \cite{Charmousis}). Here, we show that a DGP-like model can also be generated for co-dimension 1 branes at leading order in curvatures.

The Einstein Gauss-Bonnet theory in five dimensions is
\be
S_g=\frac{M_p^3}{2}\int d^4x d\chi\sqrt{-g}\left[R+\frac{1}{M_g^2}GB\right],\nonumber
\ee
where $M_g$ is a mass scale and
$GB=R_{ABCD}R^{ABCD}-4R_{AB}R^{AB}+R^2$.

The idea now is to consider an expansion in curvatures. In this case, the lowest order gravity theory will be
\be\label{gravity}
\!\!\!\!\!\!\!\!\!S_g\simeq \frac{M_p^3}{2}\int d^4x d\chi\sqrt{-g}\left[\delta R+\frac{1}{M_g^2} B_{ABCD}\delta R^{ABCD}\right],
\ee
where $\delta R_{ABCD}$ is the curvature fluctuation around the background and
$B_{ABCD}=R_{ABCD}-4 R_{BC}g_{AD}+Rg_{BC}g_{AD}$.
With the ansatz (\ref{metric}) we obtain
\be
B_{ABCD}&=&-\frac{3}{l^2}\left(g_{AC}g_{BD}-g_{BC}g_{AD}\right)+\cr
&+&\frac{4}{l}\delta(\chi)\left(g_{\alpha\gamma}-g_{\beta\gamma}g_{\alpha\delta}\right)\delta^\alpha{}_A\delta^\beta{}_B\delta^\delta{}_D\delta^\gamma{}_C .
\ee
Plugging back to the action (\ref{gravity}) we get
\be
S_g&\simeq& \frac{M_p^3}{2}\int d^4x d\chi\sqrt{-g}\Big[(1-\frac{6}{l^2M_g^2})\delta R+\cr
&+&\frac{8\delta(\chi)}{l M_g^2}\left(\delta {\cal R}+K_{\mu\nu}K^{\mu\nu}-(K^\mu{}_\mu)^2\right)\Big],
\ee
where $\delta {\cal R}$ is the induced four dimensional curvature of the brane and $K_{\mu\nu}\equiv-\frac{1}{2}\partial_\chi \gamma_{\mu\nu}$ is the extrinsic curvature of the four dimensional metric fluctuations. Again, we see that there is an instability if the bulk background curvature is such that $l<l_c$ where $l_c\equiv\sqrt{6}/M_g$ and a strong coupling in the tuned case $l=l_c$. We can now identify the effective five and four dimensional Planck masses as
$ M_5=M_p\left(1-\frac{l_c^2}{l^2}\right)^{1/3}$ and $M_4=2\sqrt{2}\frac{M_p^{3/2}}{\sqrt{l} M_g}$.
Note that if $l\sim l_c$ and all SM fields live on the brane, gravity propagates mainly in four-dimensions and $K_{\mu\nu}\rightarrow 0$, i.e. gravity is four-dimensional in this limit. This fact is in agreement with earlier studies of Newton laws for Gauss-Bonnet braneworlds \cite{sa}. With this definitions we have 
\be\label{DGP}
S_g&\simeq& \frac{1}{2}\int d^4x d\chi\sqrt{-g}\Big[M_5^3\delta R+\cr
&+&M_4^2\delta(\chi)\left(\delta {\cal R}+K_{\mu\nu}K^{\mu\nu}-K^2\right)\Big].
\ee
The action (\ref{DGP}) is not quite the DGP action. In fact it differs from it by the extrinsic curvature contribution. The DGP model suffers, in specific backgrounds, of ghost instability and/or strong coupling \cite{ghost}. Since our theory comes from a well defined purely five dimensional action, we expect that the ghost appearing in the standard DGP action will be removed by properly taking into account the extrinsic curvature contribution. We leave this important check for future work.

\section{Conclusions}

In their simplicity, braneworld scenarios, traditionally assume a special role for gravity. There, gravity is supposed to propagate everywhere so to become weaker for a brane observer, whereas the standard model of particle physics (or extensions of it), at least the gauge sectors, would only be allowed to live on a brane. Of course this hypothesis is not satisfactory from an effective field theory point of view. Moreover, requirement of diffeomorphism invariance clashes with these assumptions.

Physical branes are {\it not} lower dimensional hypersurface but domain walls. Domain walls are self-gravitating solutions generated by a field smoothly connecting two vacuums, in a short distance $L$.  Therefore, at energy scales larger than $1/L$, an observer would see a higher dimensional object. In this sense then a theory of dynamical localization of the SM urges. 

In this paper we propose a gravitational mechanism such that the kinetic term of matter fields is spontaneously localized on a domain wall, so to reproduce the (quasi-)localization hipotesys of \cite{dg}. Our mechanism is very easy to understand. A self-gravitating domain wall solution produces a peak of large curvatures at the domain wall position. By non-minimally coupling the kinetic term of SM fields to curvatures, we can easily produce localized SM kinetic terms. This localization makes the SM fields looking lower dimensional for a brane observer, similarly as in the Dvali-Shifman mechanism \cite{ds}. The only requirement we adopted for these curvature interactions was to avoid generation of higher derivative theories and/or new degrees of freedom.

An important consequence of our proposal is that branes cannot be too heavy as they would be destroyed by SM instability. 

With this mechanism at hand, one can now discuss about scatterings at energies comparable to the brane thickness, by considering specific realizations of the domain wall. At that energy scales indeed, the microstructure of the brane strongly interact, by gravity mediation, with the localized SM particles. We expect therefore new signatures from these interactions that should significantly differ from the one generated by an infinitesimally thin brane with a distributional localization of SM fields. This is of obvious importance for the ongoing LHC experiment. 

Black Holes formation in high-energy scatterings can also (in principle) be precisely addressed in our set-up.
Related to that, in our theory, the SM ``thickness'' does not generically follow the equivalence principle. In other words gravitational signatures at high energies via SM collisions would produce specific signatures dependently on the spin. We leave however the phenomenological study of this theory for future work.

We finally conclude by noticing that self-gravity interactions via the bulk Gauss-Bonnet term produces a theory of gravity with induced curvatures on the brane resembling the DGP model. This theory, however, differs form it by the presence of localized extrinsic curvatures with respect to the brane (at least for branes co-dimensions $<4$ \cite{Charmousis}). We believe that this modification should eliminate the problem of ghost propagation \cite{ghost} in cosmological backgrounds for co-dimensions $<4$ branes. This is also left for a future investigation.

\paragraph{Acknowledgments}
I would like to thank Gia Dvali for making me interested in the problem of gauge localization on a brane. I also like to thank Florian K\"unnel and Constantin Sluka for enlightening discussions, and, Gia Dvali, Stefan Hofmann, Alex Kehagias, Parvin Moyassari and Alex Pritzel for important comments on the first draft of this paper. I am supported by Alexander Von Humboldt Foundation.

\end{document}